\documentclass[a4paper]{article}

\usepackage{cite}
\usepackage{algorithm}
\usepackage{algorithmic}
\ifx\pdfoutput\undefined
\usepackage{graphicx}
\else
\usepackage[pdftex]{graphicx}
\fi
\usepackage{subfigure}
\usepackage{array}

\begin{document}

\title{A Fast Block Matching Algorithm for Video Motion Estimation Based on
Particle Swarm Optimization and Motion Prejudgment}


\author{Ran Ren\thanks{School of Communications and Information Engineering,
Nanjing University of Posts and Telecommunications. P.O. Box 214,
No.66 Xinmofan Road, Nanjing, China 210003.
Email: ranren@acm.org}\\
Madan mohan Manokar\thanks{Digibee Microsystems, Chennai, India,
600004 madanmohan@dgbmicro.com}\\
Yaogang Shi\thanks{School of Communications and Information
Engineering, Nanjing University of Posts and Telecommunications.
P.O. Box 214, No.66 Xinmofan Road, Nanjing, China 210003. Email:
madanmohan@dgbmicro.com}\\
Baoyu Zheng\thanks{Institute for Signal and Information
Processing, Nanjing University of Posts and Telecommunications.
P.O. Box 214, Nanjing, China 210003. Email: zby@njupt.edu.cn}}
\maketitle

\begin{abstract}
In this paper, we propose a fast 2-D block-based motion estimation
algorithm called \textit{Particle Swarm Optimization - Zero-motion
Prejudgment}(PSO-ZMP) which consists of three sequential routines:
1)Zero-motion prejudgment. The routine aims at finding static
macroblocks(MB) which do not need to perform remaining search thus
reduces the computational cost; 2)Predictive image coding and
3)PSO matching routine. Simulation results obtained show that the
proposed PSO-ZMP algorithm achieves over 10 times of computation
less than Diamond Search(DS) and 5 times less than the recent
proposed Adaptive Rood Pattern Searching(ARPS). Meanwhile the PSNR
performances using PSO-ZMP are very close to that using DS and
ARPS in some less-motioned sequences. While in some sequences
containing dense and complex motion contents, the PSNR
performances of PSO-ZMP are several dB lower than that using DS
and ARPS but in an acceptable degree.
\end{abstract}


\section{Introduction}
With the increasing popularity of technologies such as digital
television, Internet streaming video and video conferencing, video
compression has became an essential component of broadcast and
entertainment media. Among various kinds of approaches,
block-based motion estimation and compression are most widely
accepted ones. The \textit{block-matching algorithm}(BMA) for
\textit{motion estimation}(ME) has been adopted in many
international standards for digital video compression, such as
H.264 and MPEG 4\cite{video.compress.book}. In the framework of
video coding, the statistical redundancies can be categorized by
either temporal or spatial. For the purpose of reducing the
temporal redundancies among frames, motion estimation was
applied\cite{me.digital.tv}. Block-based matching algorithms
consider each frame in the video sequence formed by many
nonoverlapping small regions, called the marcoblocks(MB) which are
often square-shaped and with fixed-size($16\times 16$ or $8\times
8$). Let $\mathcal{B}_{m}$ represents the $m$th MB and
$\mathcal{M}$ the number of blocks, and $\mathcal{M} =
{1,2,\cdots, M}$; let $\Lambda$ be the entire frame and the
partition into MBs should satisfy $\bigcup \mathcal{B}_{m} =
\Lambda$ and $\mathcal{B}_{m} \bigcap \mathcal{B}_{n} = \O, m \neq
n$\cite{video.proc.comm}. Given a MB $\mathcal{B}_{m}$ in the
anchor frame, the motion estimation problem is to determine a
corresponding matching MB $\mathcal{B}_{m}'$ in the target frame
such that the matching error between these two blocks is
minimized. Then, a motion vector is computed by subtracting the
coordinates of the MB in the anchor frame from that of the
matching MB in the target frame. Instead of sending the entire
frame pixel-by-pixel, a set of motion vectors is transmitted
through the channel which greatly reduces the amount of
transmission. In the decoder side, a motion compensated procedure
is applied to reconstruct frames using the received motion vectors
and the anchor frame. Referred to many researches, the motion
estimation and encoding part consumes nearly $70-90$ percent of
the total amount of computation in the whole video compression
procedure thus making it an active research topic in the last two
decades.

There are many proposals of BMAs in literature. The most basic one
is the \textit{Exhaustive Search}(ES), also known as full search
which simply compares the given MB in the anchor frame with all
candidate MBs in the target frame exhaustively within a predefined
search region. Previous research showed that ES can obtain high
matching accuracy but requires a very large amount of computation
thus infeasible to implement in real-time video applications. To
speed up the search, various fast algorithms for block matching
which reduce the number of search candidates have been developed.
Well known examples are \textit{2-D Logarithmic
Search}(LOGS)\cite{logs}, \textit{Three Step
Search}(TSS)\cite{tss}, \textit{Four Step Search}(4SS)\cite{4ss},
\textit{Diamond Search}(DS)\cite{ds} which is accepted in the
MPEG-4 Verification Model and widely implemented in VLSI, and the
recent proposed \textit{Adaptive Rood Pattern
Search}(ARPS)\cite{arps} which is almost two or three times faster
than DS and even achieves higher \textit{peak signal-to-noise
ratio}(PSNR) than that using DS.

From the optimization point of view, block-based methods can be
described by the following minimization\cite{handbook.video.proc},
$\forall \mathbf{m}$:

\begin{displaymath}
 \min_{\mathbf{d}_{m}\in
\mathcal{P}}\varepsilon(\mathbf{d}_{m}), \
\varepsilon(\mathbf{d}_{m}) = \sum_{\mathbf{n}\in \mathcal{B}_{m}}
\Phi(I_{k}[\mathbf{n}] - I_{k-1}[\mathbf{n}+\mathbf{d}_{m}])
\end{displaymath}

where $I_k$ is the target frame; $I_{k-1}$ is the anchor frame;
$\varepsilon(\mathbf{d}_{m})$ is the matching error; $\mathbf{d}$
are the motion vectors and $\mathcal{P}$ is the search area to
which $\mathbf{d}_{m}$ belongs, defined as $\mathcal{P} =
{\mathbf{n}=(n_1,n_2):-P \leq n_1 \leq P, -P \leq n_2 \leq P}$.
Sign of $d$ is positive when motion of the block is towards
positive direction from $k-1$th frame to $k$th frame. And negative
when motion of the blcok is in negative direction from $k-1$th
frame to $k$th frame. $\mathcal{B}_m$ is an $N \times N$ size MB
with the top-left corner coordinate at $\mathbf{m} = (m_1,m_2)$.
The goal is to find the best displacement motion vector
$\mathbf{d}_m$ for each MB $\mathcal{B}_m$, in the sense of the
criterion $\Phi$.

\textit{Particle swarm optimization}(PSO) was originally proposed
by Kennedy and Eberhart in 1995\cite{pso.original}. It is widely
accepted and focused by researchers due to its profound
intelligence background and simple algorithm structure. Currently,
PSO has been implemented in a wide range of research areas such as
functional optimization, pattern recognition, neural network
training, fuzzy system control etc. and obtained significant
success. Like \textit{Genetic Algorithm}(GA), PSO is also an
evolutionary algorithm based on swarm intelligence. But, on the
other side, unlike GA, PSO has no evolution operators such as
crossover and mutation\cite{pso.ga.comp}. In PSO, the potential
solutions, called particles, fly through the solution space by
following the current optimum particles. The original intent was
to graphically simulate the graceful but unpredictable
choreography of a bird flock. Through competitions and
cooperations, particles follow the optimum points in the solution
space to optimize the problem. Many proposals indicate that PSO is
relatively more capable for global exploration and converges more
quickly than many other heuristic algorithms\cite{pso.res}.

The rest of the paper is organized as follows. Section II
introduces the PSO algorithm and we propose the PSO-ZMP
block-matching algorithm for motion estimation in Section III.
Simulation results and analysis on five video sequences are given
in Section IV. Section V concludes the paper.

\section{Particle Swarm Optimization}
Particle swarm algorithm is a kind of evolutionary algorithm based
on swarm intelligence. Each potential solution is considered as
one particle, and these particles are distributed stochastically
in the high-dimensional solution space in the initialization
period of the algorithm. Through following the optimum discovered
by itself and the entire group, each particle periodically updates
its own velocity and position.

\setlength{\arraycolsep}{0.0em}
\begin{eqnarray}
    v_{id}(t+1)&{} ={}& w \times v_{id}(t) + c_{1} \times rand_{1}(\cdot)\nonumber\\
    &&\times(p_{id}-x_{id}) + c_{2} \times rand_{2}(\cdot) \nonumber\\
    &&\times(p_{gd}-x_{id})
    \label{vid.original}
\end{eqnarray}
\setlength{\arraycolsep}{5pt}

\begin{equation}
x_{id}(t+1) = x_{id}(t) + v_{id}(t+1) %
\label{xid.original} %
\end{equation}

$$
1\leq i \leq N, 1\leq d \leq D %
$$

Where, $N$ is the number of particles and $D$ is the
dimensionality; $\mathbf{V}_{i} = (v_{i1},v_{i2},\cdots, v_{iD})$,
$v_{id}\in [-v_{max}, v_{max}]$ is the velocity vector of particle
$i$ which decides the particle's displacement in each iteration.
Similarly, $\mathbf{X}_{i} = (x_{i1},x_{i2},\cdots, x_{iD})$,
$x_{id}\in [-x_{max},x_{max}]$ is the position vector of particle
$i$ which is a potential solution in the solution space. the
quality of the solution is measured by a fitness function; $w$ is
the inertia weight which decreases linearly during a run; $c_{1},
c_{2}$ are both positive constants, called the acceleration
factors which are generally set to 2.0; $rand_1(\cdot)$ and
$rand_2(\cdot)$ are two independent random number distributed
uniformly over the range $[0,1]$; and $p_{g}$, $p_{i}$ are the
best solutions discovered so far by the group and itself
respectively.

In the $t+1$ time iteration, particle $i$ uses $p_{g}$ and $p_{i}$
as the heuristic information to updates its own velocity and
position. The first term in Eq.\ref{vid.original} represents the
diversification, while the second and third intensification. The
second and third terms should be understood as the trustworthiness
towards itself and the entire social system respectively.
Therefore, a balance between the diversification and
intensification is achieved based on which the optimization
progress is possible.

\section{Block-matching algorithm based on PSO-ZMP}
In this paper, an algorithm based on \textit{Particle Swarm
Optimization}(PSO) and \textit{Zero-Motion Prejudgment}(ZMP) is
proposed to reduce the computation and obtain satisfied
compensated video quality. The PSO-ZMP algorithm consists of three
sequential routines. 1)Zero-motion prejudgment; 2)Predictive image
coding; 3)PSO matching. Instead of distributed stochastically in
the entire matching space, we also devise a novel distribution
pattern for particle initialization to bear the center-biased
characteristics of common motion fields.

\subsection{Performance Evaluation Criterion}
As widely adopted, we measure the amount of computation and the
quality of compensated video sequence by \textit{Computation} and
\textit{Peak Signal-to-Noise Ratio}(PSNR). Computation is defined
as the average number of the error function evaluations per MV
generation. Due to the minimum computational cost, we choose
\textit{Summed Absolute Difference}(SAD) as the error function
which is defined as follows:

\begin{equation}
SAD = \frac{1}{N}
\sum_{i=1}^{N}\sum_{j=1}^{N}(|I_k(i,j)-I_{k-1}(i,j)|) %
\label{eq.SAD}
\end{equation}

where the size of a MB is $N \times N$.\\

The motion estimate quality between the original $I_{ogn}$ and the
compensated video sequences $I_{cmp}$ is measured in PSNR which is
defined as:

\begin{displaymath}
 PSNR = 10\log_{10} \frac{I_{\max}^2}{\sigma_{e}^2}
\end{displaymath}

\begin{displaymath}
\sigma_{e}^2 = MSE =
\frac{1}{N}\sum_{k=0}^K\sum_{i=0}^{N}\sum_{j=0}^{N}(I_{ogn}(i,j,k)-I_{cmp}(i,j,k))^2
\end{displaymath}

where $K$ is the number of frames in the video sequence.

\subsection{Zero-Motion Prejudgment}
Zero-Motion Prejudgment(ZMP) was firstly introduced in
\cite{arps}. Data shown in \cite{arps} represented that in most of
test sequences, more than 70\% of the MBs are static which do not
need the remaining search. So, significant reduction of
computation is possible if we perform the ZMP procedure before the
follow-up predictive coding and PSO matching routine. We first
calculate the matching error(SAD in this paper) between the MB in
the anchor frame and the MB at the same location in the target
frame and then compare it to a predetermined threshold, saying
$\Delta$. If the matching error is smaller than $\Delta$, we
consider this MB static which do not need any further motion
estimation, and return a $[0, 0]$ as its motion vector(MV).

\subsection{Predictive Image Coding}
Based on the center-biased characteristics in video sequences,
that is, certain MBs are highly correlated in local regions of the
frame, the encoder creates a prediction of a region of the current
frame based on previously encoded and transmitted frames.

\begin{figure}
\centering %
\includegraphics[width=2.0in, height = 2.0in]{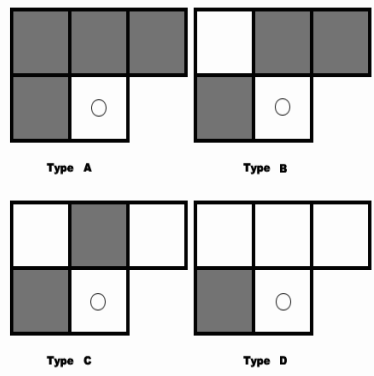} %
\caption{Four types of ROS for current-encoded MB. (The block
marked ``$\bigcirc$" is the current-encoded MB; Blocks in grey are
the reference MBs for prediction.)} \label{pre.pattern}
\end{figure}

If the frame is processed in raster order, the current-encoded MB
should have four patterns of \textit{region of support}(ROS) that
consists of the neighboring blocks whose MVs will be used to
compute the predicted MV for prediction in Fig.~\ref{pre.pattern}
due to the limited computational cost. Experiment mentioned in
\cite{arps} shows there is little PSNR difference using these four
ROS patterns in the predictive coding routine, and ROS type D
consumes least amount of computation because of its simplest
structure. Thus, ROS pattern D is adopted in this paper.


\begin{figure}
\centering %
\includegraphics[width=2.0in, height = 2.0in]{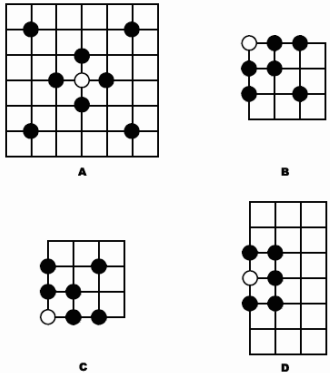} %
\caption{Patterns to initial particles} \label{pattern.all}
\end{figure}

\subsection{Selection of Search Patterns}
Due to the spatial correlation characteristics between MBs in one
frame, during the initiation period of the PSO matching routine,
we distribute the particles in four specific
patterns(Fig.~\ref{pattern.all}) with a view to reduce the
computational cost but to achieve higher PSNR.

Since frames are processed in raster order, the MB in the top-left
corner in the frame, can not be predictive coded because there is
no reference MB for prediction in the current-encoded frame. Thus,
for this condition, we simply skip the predictive coding and begin
PSO searching routine directly with the initial positions of
particles in the pattern type B in Fig.~\ref{pattern.all}.

For those MBs located at the leftmost column of frames, their
reference MBs used in predictive coding are in the other side of
the frame, thus may not be highly correlated and inefficient in
prediction. So, we also solely perform the PSO searching routine
in this case, with the pattern type D in Fig.~\ref{pattern.all}.
And, for the last leftmost MB been processed in the frame, that
is, the MB in the bottom-left corner, we use the pattern type C in
Fig.~\ref{pattern.all} instead.

Otherwise, pattern type A in Fig.~\ref{pattern.all} is adopted. We
put four particles in a rood shape with size zero(size refers to
the distance between any vertex point and the center-point) in the
adjacent MBs and four particles in a rood shape with size one, and
then rotate it by angle $\pi /2$. With two rood shape in
difference size, we try to balance the global exploration and
local refined search in order for broader searching space as well
as higher matching accuracy. Moreover, we distribute particles
equally in all directions(8 particles in 8 directions) with a view
to, in stochastic condition, find the matching MB in each
direction with equal possibility.

Notably, if the position of a particle in the during
initialization and a PSO run is out of the boundary of the image
frame, we simply put the particle in the position nearest to its
intended position.

\subsection{Stopping Criterion}
Generally, there are two widely adopted stopping criteria. One is
\textit{Fixed-iteration}, that is, given a certain iteration time,
saying $N$, the search stops after $N$ times of iteration. The
other is \textit{Specified-threshold}. During a PSO run, the
most-fitted value found by the entire group $p_g$, called the
``\textit{best so far}" value will be updated by the particles.
For minimization problems, we specify a very small threshold
$\varepsilon$, and if the change of $p_g$ during $t$ times of
iteration is smaller than the threshold, we consider the group
best value very near to the global optimum, thus the matching
procedure stops. Due to the center-biased characteristics of
real-world motion fields, we adopt the fixed-iteration method in
this paper for reducing the computational cost.

\subsection{Summary of Our Method}
We incorporate the ZMP, the predictive coding and the PSO matching
routines together and propose a block-matching algorithm for
motion estimation based on PSO and
ZMP. The algorithm can be summarized in the pseudocode below:\\

\begin{algorithm}
\caption{PSO-ZMP BMA} \label{pseudo.pso.bma}
\begin{algorithmic}[1]
\FOR{Each frame $i$}%
    \FOR{Each MB $j$}%
        \STATE{$zmpcost \leftarrow SAD(I_{i-1}(j), I_i(j))$} %
        \IF{$zmpcost < \mathcal{\Delta}$} %
            \STATE{Consider MB $j$ static} %
            \STATE{motionVect $ = [0,0]$} %
            \STATE{Continue} %
        \ELSE %
            \IF{MB $j$ in the leftmost column of frame $i$}%
                \IF{MB $j == 1$} %
                    \STATE{Initial particles in pattern B, Fig.~\ref{pattern.all}} %
                \ELSIF{MB $j$ in the bottomright corner of frame $i$} %
                    \STATE{Initial particles in pattern C, Fig.~\ref{pattern.all}} %
                \ELSE %
                    \STATE{Initial particles in pattern D, Fig.~\ref{pattern.all}} %
                \ENDIF
            \ELSE %
                \STATE{Initial particles in pattern A, Fig.~\ref{pattern.all}} %
                \STATE{Predictive image coding routine} %
            \ENDIF %
            \STATE{Begin PSO matching routine}%
            \FOR{Each iteration time $t$} %
                \FOR{Each particle $p$} %
                    \STATE{Evaluate $SAD$ using Eq.~\ref{eq.SAD} and update $P_g$,
                        $P_p$}
                    \STATE{Update velocity using Eq.~\ref{vid.original}} %
                    \STATE{Update position using Eq.~\ref{xid.original}} %
                \ENDFOR %
            \ENDFOR %
        \ENDIF %
    \ENDFOR
    \STATE{calculate the motion vector and output}
\ENDFOR %
\end{algorithmic}
\end{algorithm}

\section{Experiments and Results}
We've tested our PSO-ZMP algorithm on five test video sequences:
Akiyo, Container, Mother \& Daughter, News and Silent within 100
image frames(except 90 frames in Akiyo due to the limitation of
the sequence length).

\subsection{Experimental Settings}
\subsubsection{PSO Parameters}
PSO matching is the core routine in our algorithm. In this paper,
to balance between computational cost and compensated video
quality, we adopt the standard PSO with inertia
weight\cite{empirical.pso, modified.pso} which is widely
considered as the defacto PSO standard. We use the fixed-iteration
stopping criterion with max 5 iterations. The max velocity is set
to 5. The inertia weight $w$ decreases linearly from 0.9 to 0.4
during a PSO run and two acceleration factors $c_1$, $c_2$ are set
to 2.0, as commonly did.

\subsubsection{Motion Estimation Parameters}
\begin{itemize}
\item We divide a whole image frame into $16 \times 16$ MBs in the simulation. %
\item We select a ZMP threshold $\Delta$ for each test video
sequence correspondingly based on data obtained in experiments.
The parameters are shown in Table~\ref{zmp.table}. %
\item We do not restrict the range of candidate matching MBs
rigidly by a search window $\mathcal{P}$. Instead, through the
fixed-iteration and the setting of max velocity,particles search
for the matching MB in an area more flexible and adaptable.
\end{itemize}

\begin{table}
\renewcommand{\arraystretch}{1.3}
\caption{ZMP threshold $\Delta$ for five test video sequences} %
\label{zmp.table}
\begin{center}
\begin{tabular}{|c|c|c|}
\hline
Sequence & Format & ZMP Threshold $\Delta$\\
\hline %
Akiyo & QCIF & 384\\
\hline
Container & QCIF & 512 \\
\hline
Mot. \& Dau. & QCIF & 384 \\
\hline
News & QCIF & 512 \\
\hline
Silent & QCIF & 384 \\
\hline
\end{tabular}
\end{center}
\end{table}

\subsection{Results and Analysis}
Fig.~\ref{fig_sim_1} and fig.~\ref{fig_sim_2} below show the
simulation results on five test video sequences. For comparison,
the performance of DS, ARPS, GA-ZMP, the BMA based on the genetic
algorithm and PSO-ZMP algorithm are examined. Average \textit{peak
signal-to-noise ratio}(PSNR) per frame of the reconstructed video
sequence is computed for quality measurement and documented in
Table~\ref{psnr.gain.table}. The computational gain of our PSO-ZMP
to DS(or ARPS) is defined by the ratio of matching speed to that
of our method, which is shown in Table~\ref{comp.gain.table}.

From the results obtained, PSO-ZMP shows significant computational
reductions while acceptable drops in \textit{peak signal-to-noise
ratio}(PSNR). Notably, in sequence Akiyo and Mother \& Daughter,
our method achieves very close PSNR performance(max difference
1.09dB in Mother \& Daughter) with 12.04 and 12.44 times of
computation reductions compared to DS respectively; 4.94 and 5.62
times of computation reductions compared to ARPS. In sequence
Silent, News and Container, the PSNR performances using our method
are 2-4 dB(max difference 4.04dB in Silent) less than that of ARPS
and DS. But, in those sequences, compared to DS, PSO-ZMP consumes
over 8-12 times less of computation to that of DS and 3-6 times
less to that of ARPS. Referred to \cite{video.proc.comm}, a PSNR
higher than 40dB typically indicates an excellent image(i.e.,
being very close to the original), between 30-40dB usually means a
good images(i.e., the distortion is visible but acceptable);
between 20-30dB PSNR is quite poor; and finally, a PSNR lower than
20dB is unacceptable. For all five sequences tested, PSO-ZMP
algorithm achieves PSNR higher than 30dB in most of the frames,
thus the PSNR droppings are in an acceptable degree.

Compared to GA-ZMP which incorporates \textit{genetic algorithm}
and \textit{zero-motion prejudgment}(ZMP), our PSO-ZMP algorithm
achieves superior performances on average PSNR and computation on
all five test sequences. With the evolution operators such as
crossover and mutation, GA consumes more amount of computation
which leads to 1.5-2 times more computations than that using PSO.
Meanwhile, our algorithm with PSO and ZMP incorporated obtains
higher average PSNR compared to that using GA because PSO is more
capable for the global exploration and local
exploitation\cite{pso.ga.comp}.

\begin{table}
\renewcommand{\arraystretch}{1.3}
\caption{Average PSNR performance of DS, ARPS and PSO-ZMP} %
\label{psnr.gain.table}
\begin{center}
\begin{tabular}{|c|c|c|c|c|}
\hline
Sequence & DS & ARPS & GA-ZMP & PSO-ZMP \\
\hline %
Akiyo & 43.50 & 43.49 & 42.07 & 42.39\\
\hline
Container & 36.34 & 36.13 & 32.36 & 33.15\\
\hline
Mot. \& Dau. & 40.46 & 40.57 & 35.66 & 39.48 \\
\hline
News & 36.66 & 36.61 & 35.02 & 35.29\\
\hline
Silent & 36.68 & 36.46 & 31.62 & 32.64\\
\hline
\end{tabular}
\end{center}
\end{table}

\begin{table*}
\renewcommand{\arraystretch}{1.3}
\caption{Computational gain to ARPS and to DS} %
\label{comp.gain.table}
\begin{center}
\begin{tabular}{|c|c|c|c|c|}
\hline
Sequence & ARPS to DS & PSO-ZMP to DS & PSO-ZMP to ARPS & PSO-ZMP to GA-ZMP\\
\hline %
Akiyo & 2.44 & 12.04 & 4.94 & 1.47\\
\hline
Container & 2.24 & 8.10 & 3.62 & 1.54\\
\hline
Mot. \& Dau. & 2.22 & 12.44 & 5.62 & 1.61\\
\hline
News & 2.32 & 9.85 & 4.25 & 1.68\\
\hline
Silent & 2.25 & 8.60 & 3.82 & 2.17\\
\hline
\end{tabular}
\end{center}
\end{table*}

%

\begin{figure*}
\centerline{\subfigure[Computations on Akiyo]{ %
\includegraphics[width=3.5in]{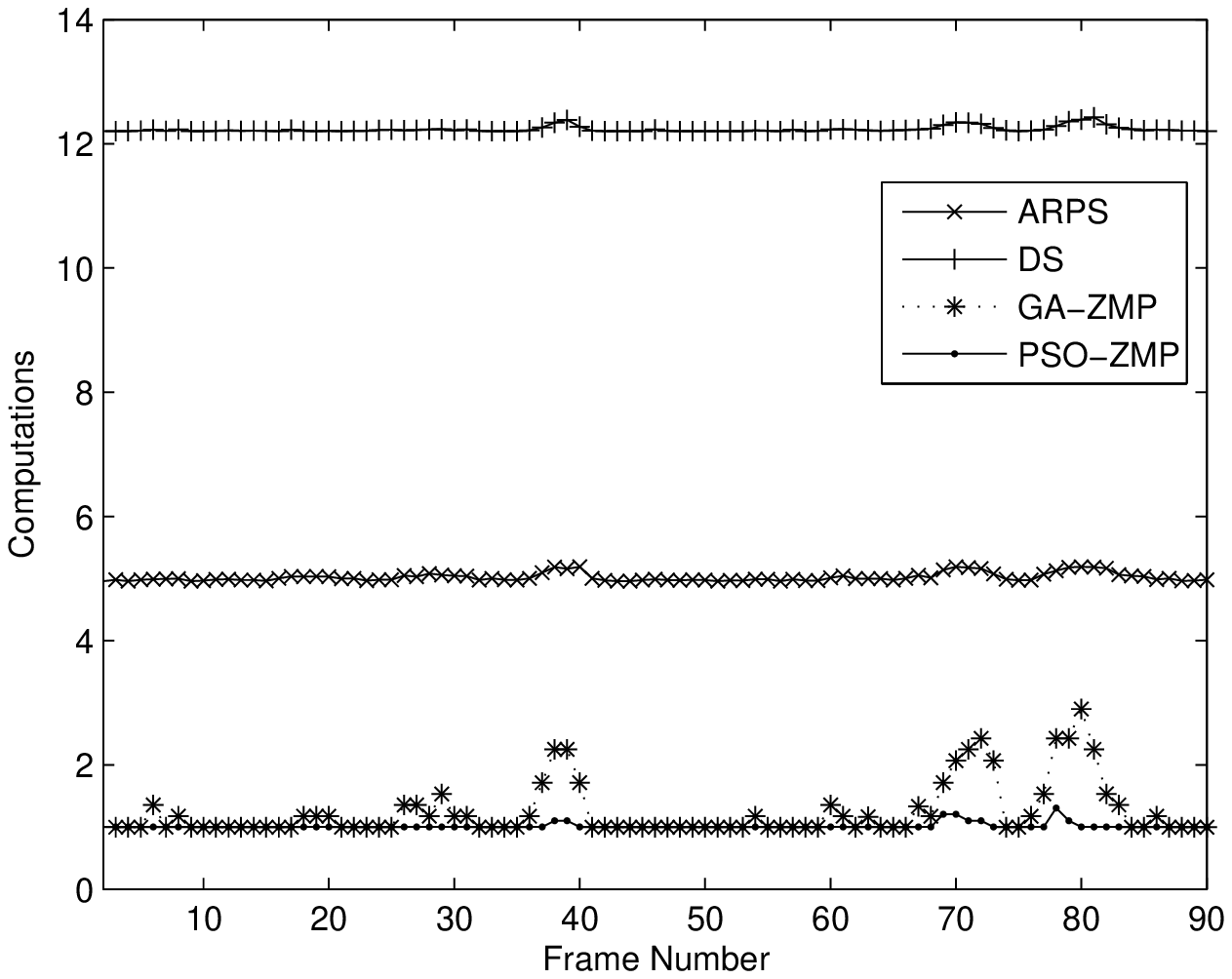}
\label{akiyo.comp}} %
\hfil %
\subfigure[PSNR on Akiyo]{ %
\includegraphics[width=3.5in]{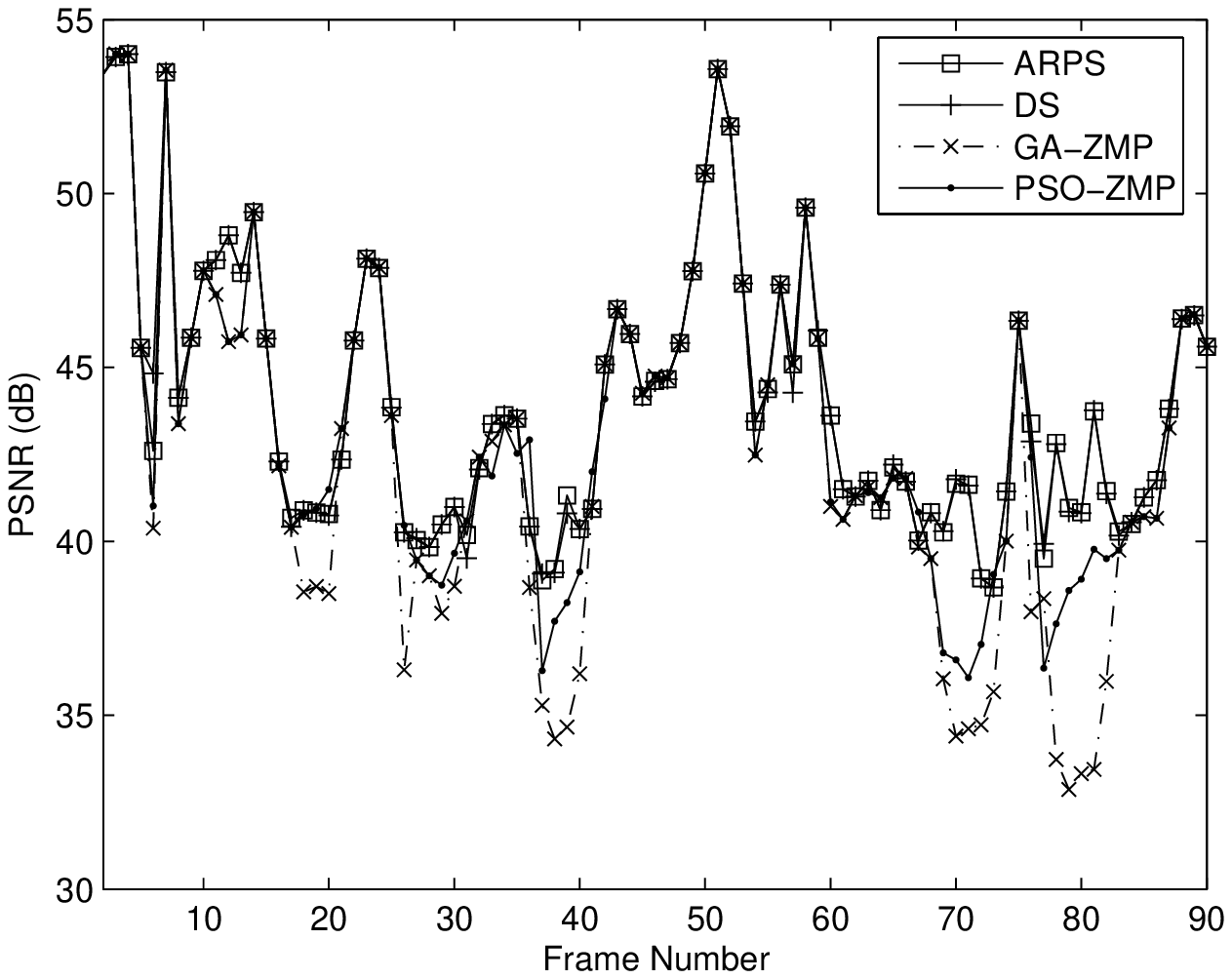}
\label{akiyo.psnr}}} %
\vfil %
\centerline{\subfigure[Computations on Container]{ %
\includegraphics[width=3.5in]{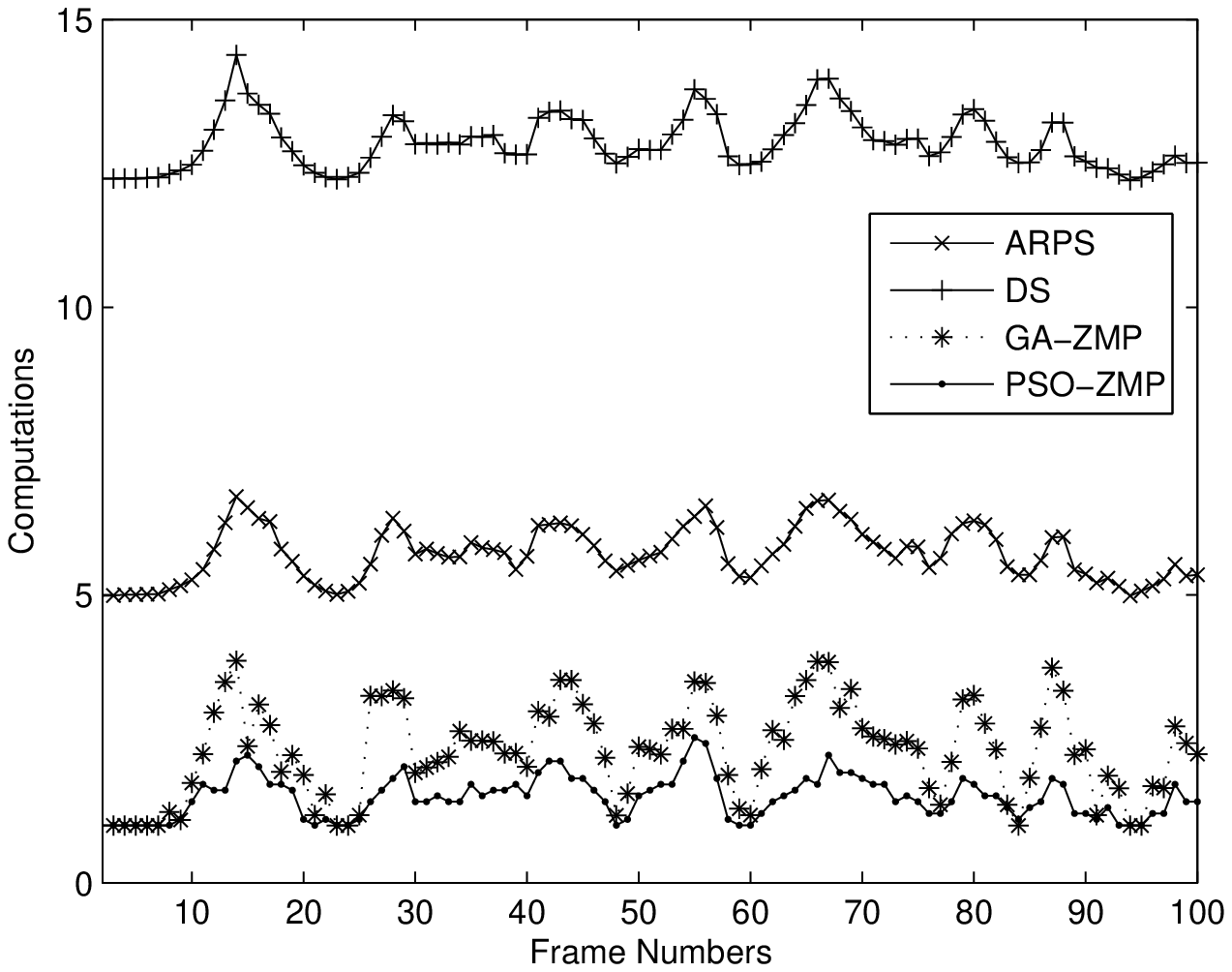}
\label{container.comp}} %
\hfil %
\subfigure[PSNR on Container]{ %
\includegraphics[width=3.5in]{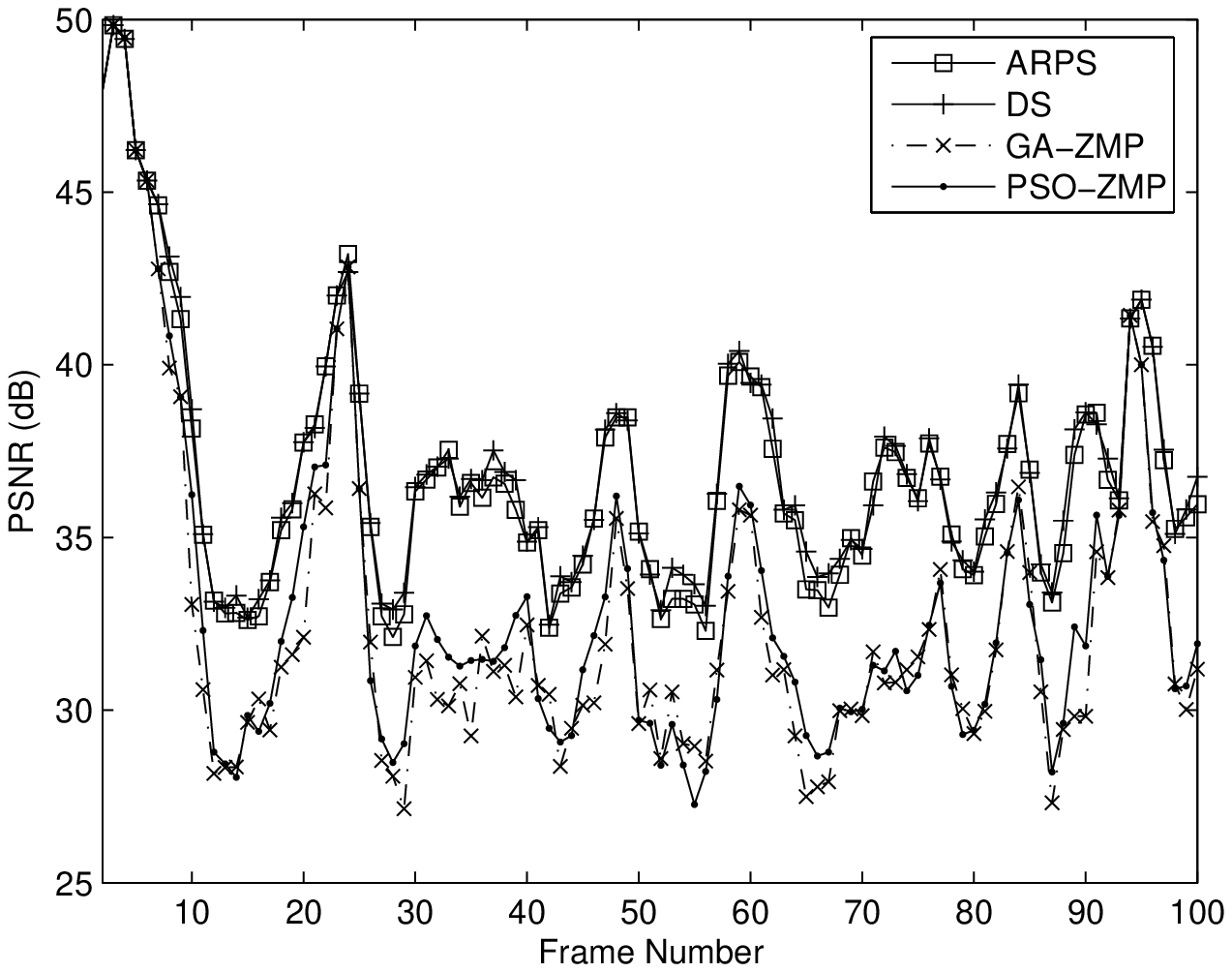}
\label{container.psnr}}} %
\vfil %
\centerline{\subfigure[Computations on Mother \&
Daughter]{ %
\includegraphics[width=3.5in]{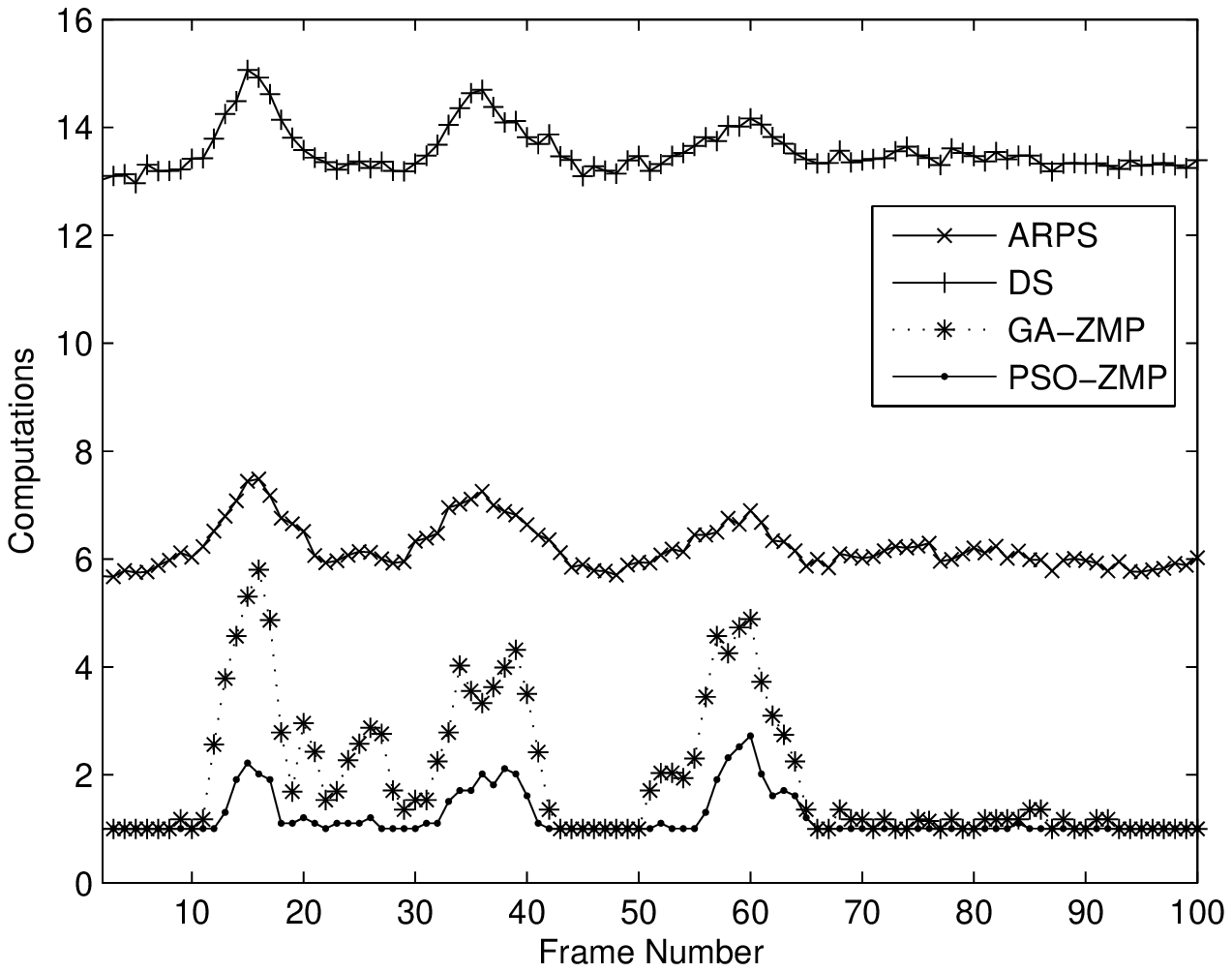}
\label{moth.dor.comp}} %
\hfil %
\subfigure[PSNR on Mather \&
Daughter]{\includegraphics[width=3.5in]{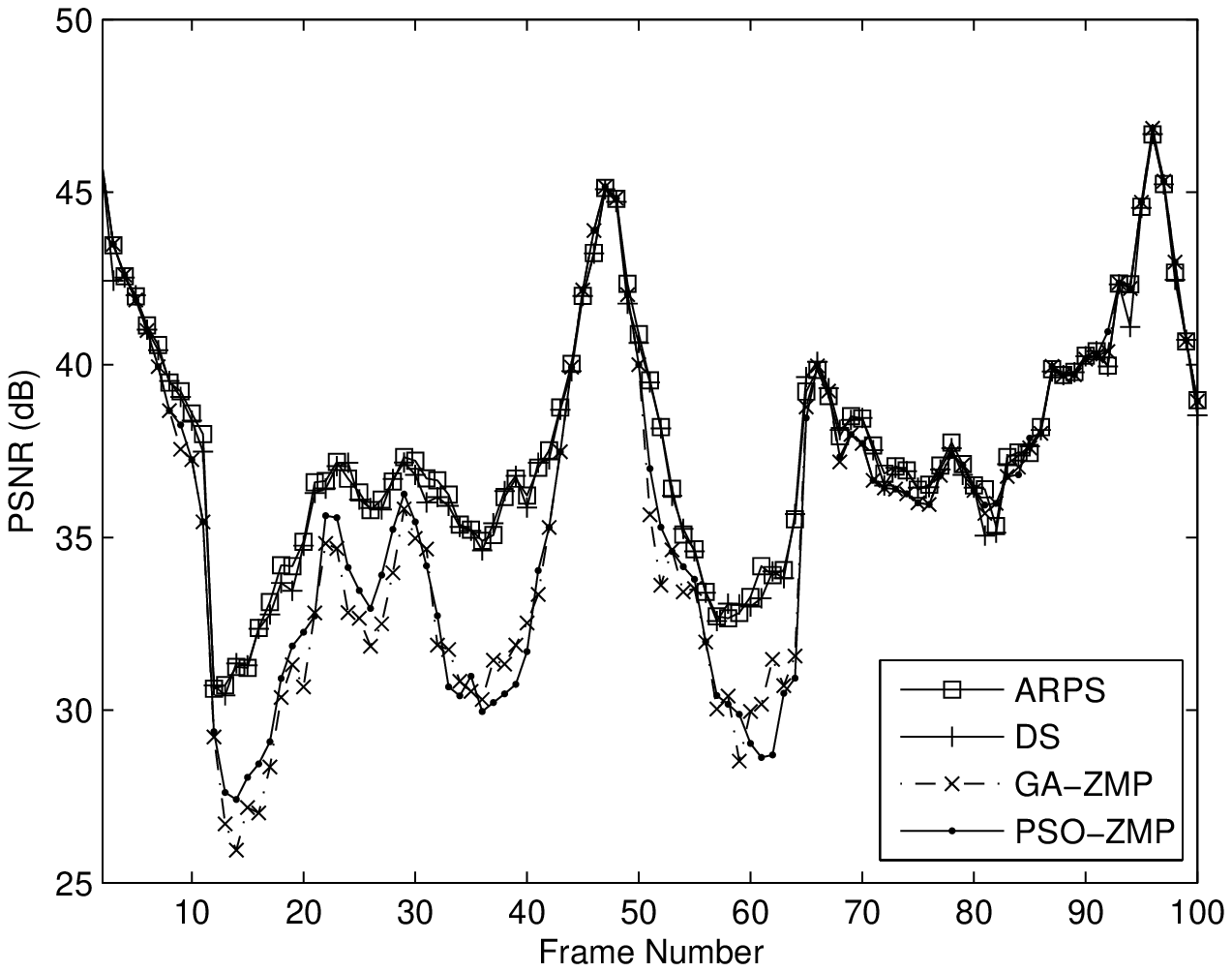}
\label{moth.dor.psnr}}} %
\caption{Simulation results on Akiyo, Container and Mother \&
Daughter} %
\label{fig_sim_1} %
\end{figure*}

\begin{figure*}
\centerline{\subfigure[Computations on
News]{\includegraphics[width=3.5in]{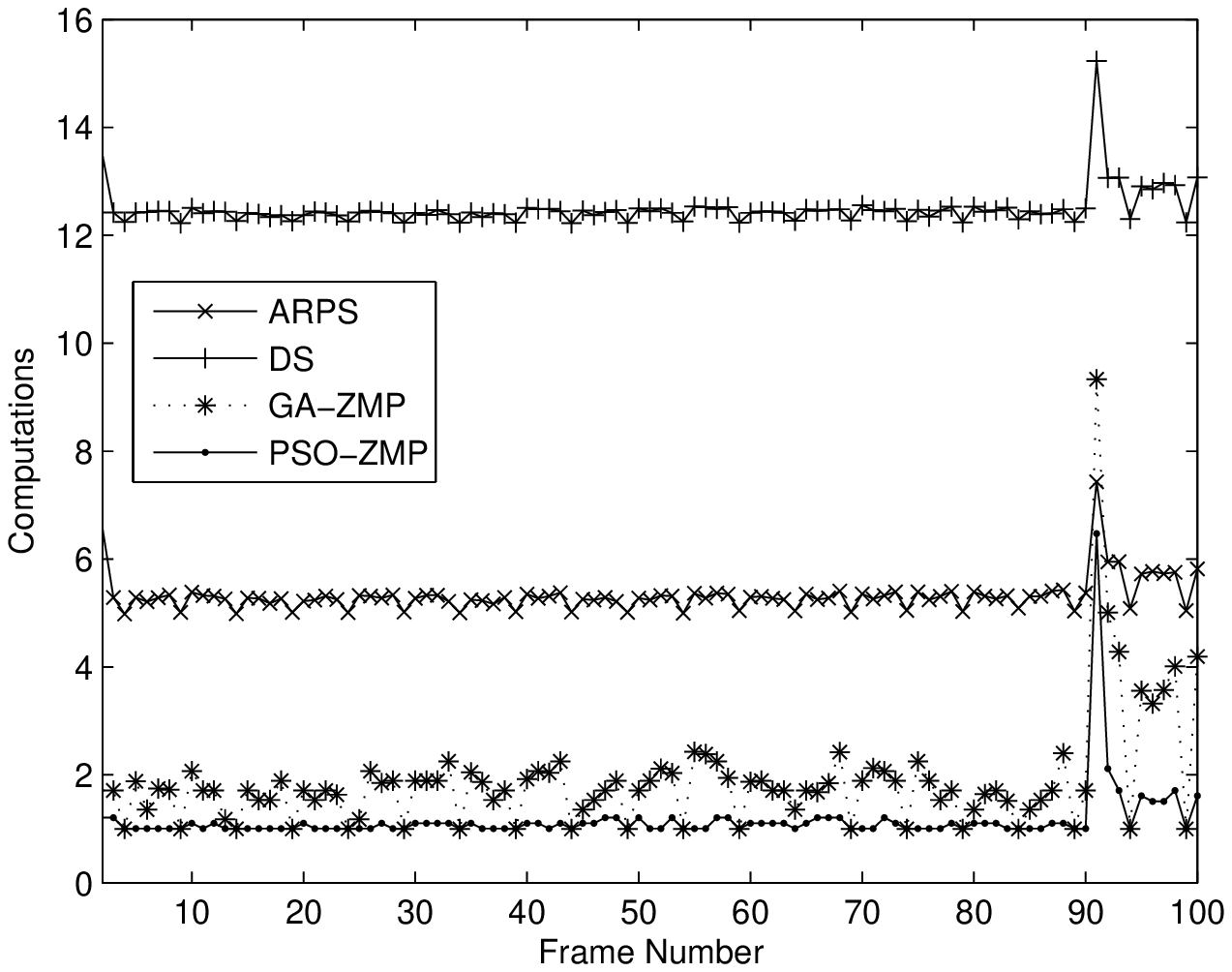}
\label{news.comp}} %
\hfil %
\subfigure[PSNR on
News]{\includegraphics[width=3.5in]{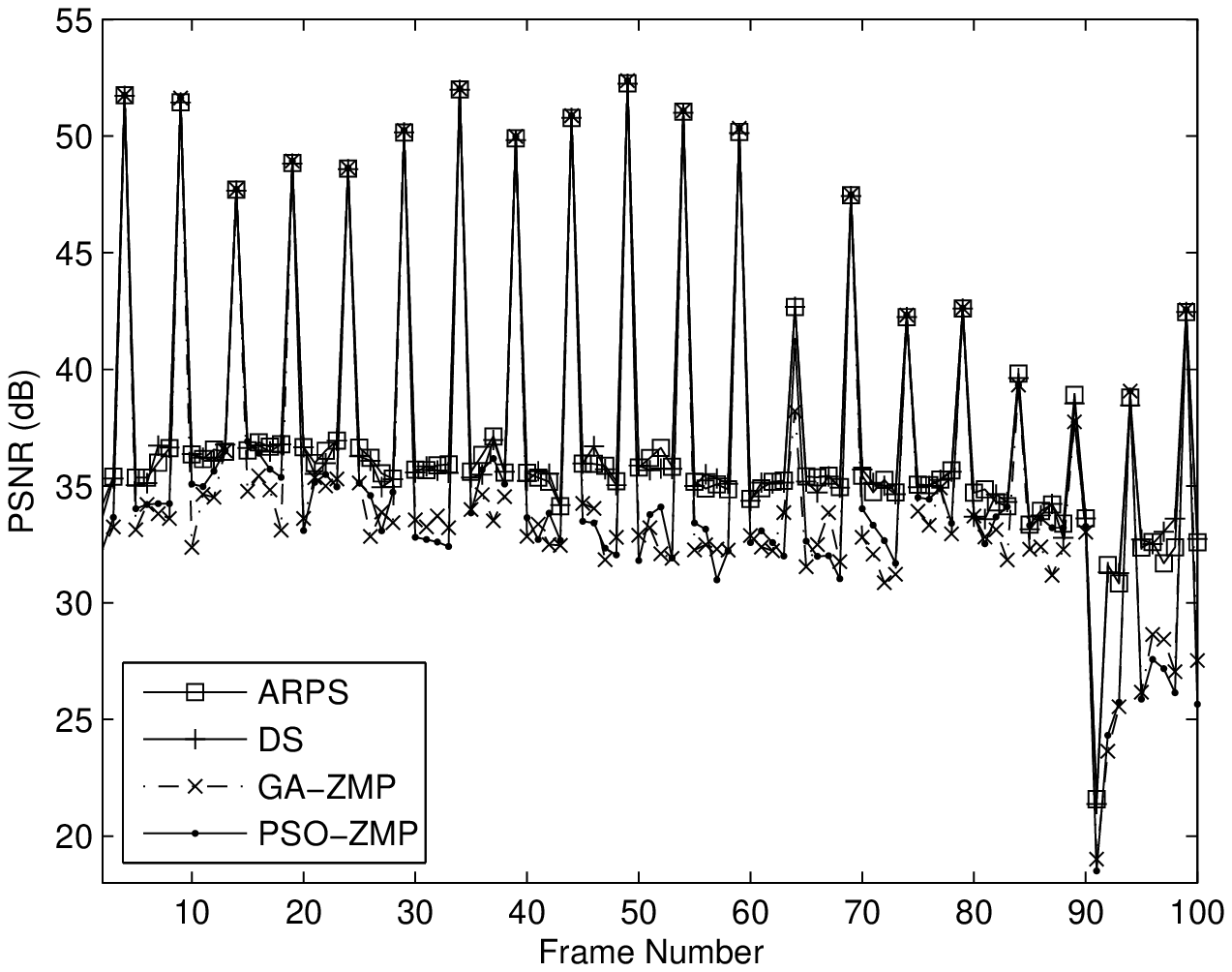}
\label{news.psnr}}} %
\vfil %
\centerline{\subfigure[Computations on
Silent]{\includegraphics[width=3.5in]{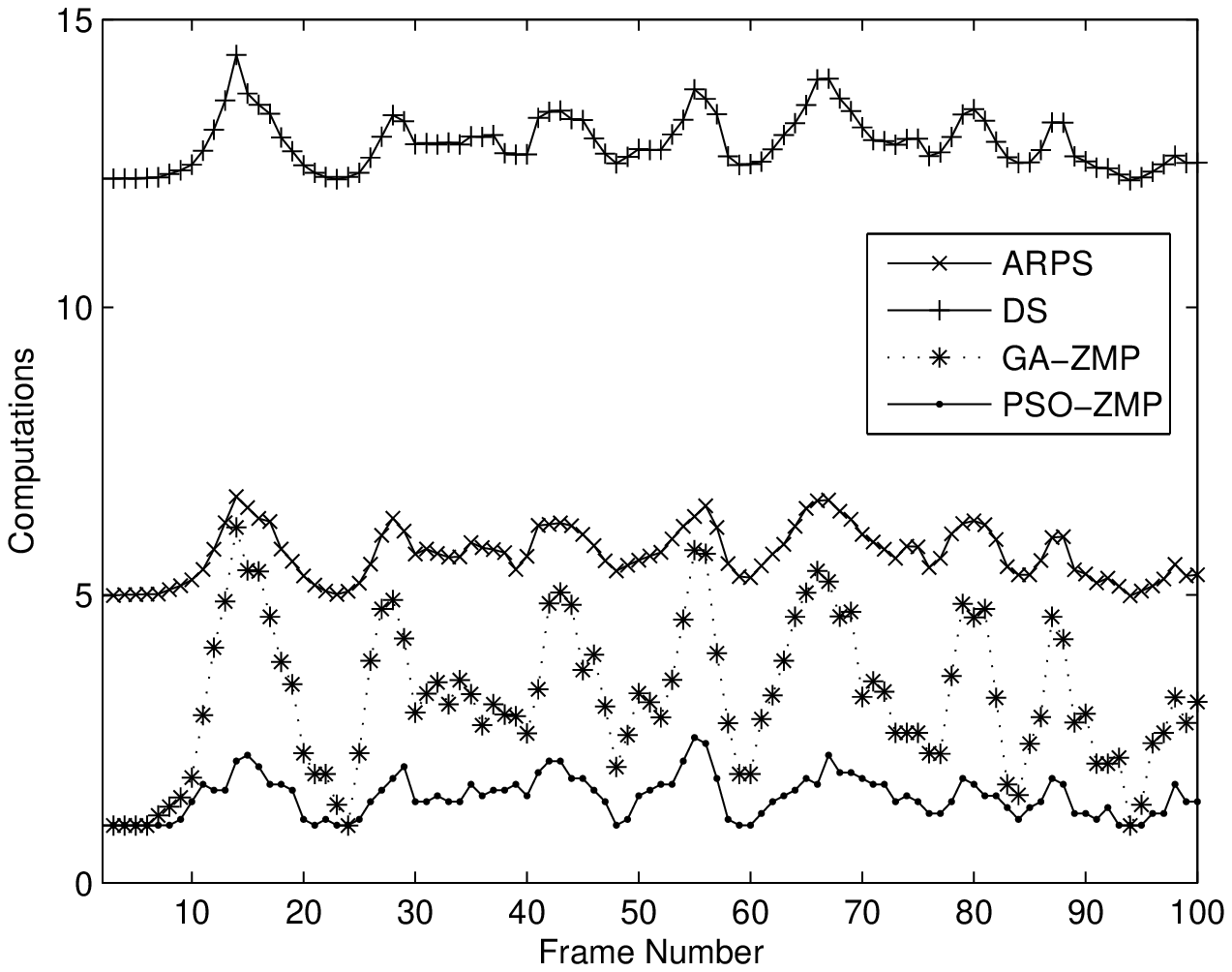}
\label{suzie.comp}} %
\hfil %
\subfigure[PSNR on Silent]{\includegraphics[width=
3.5in]{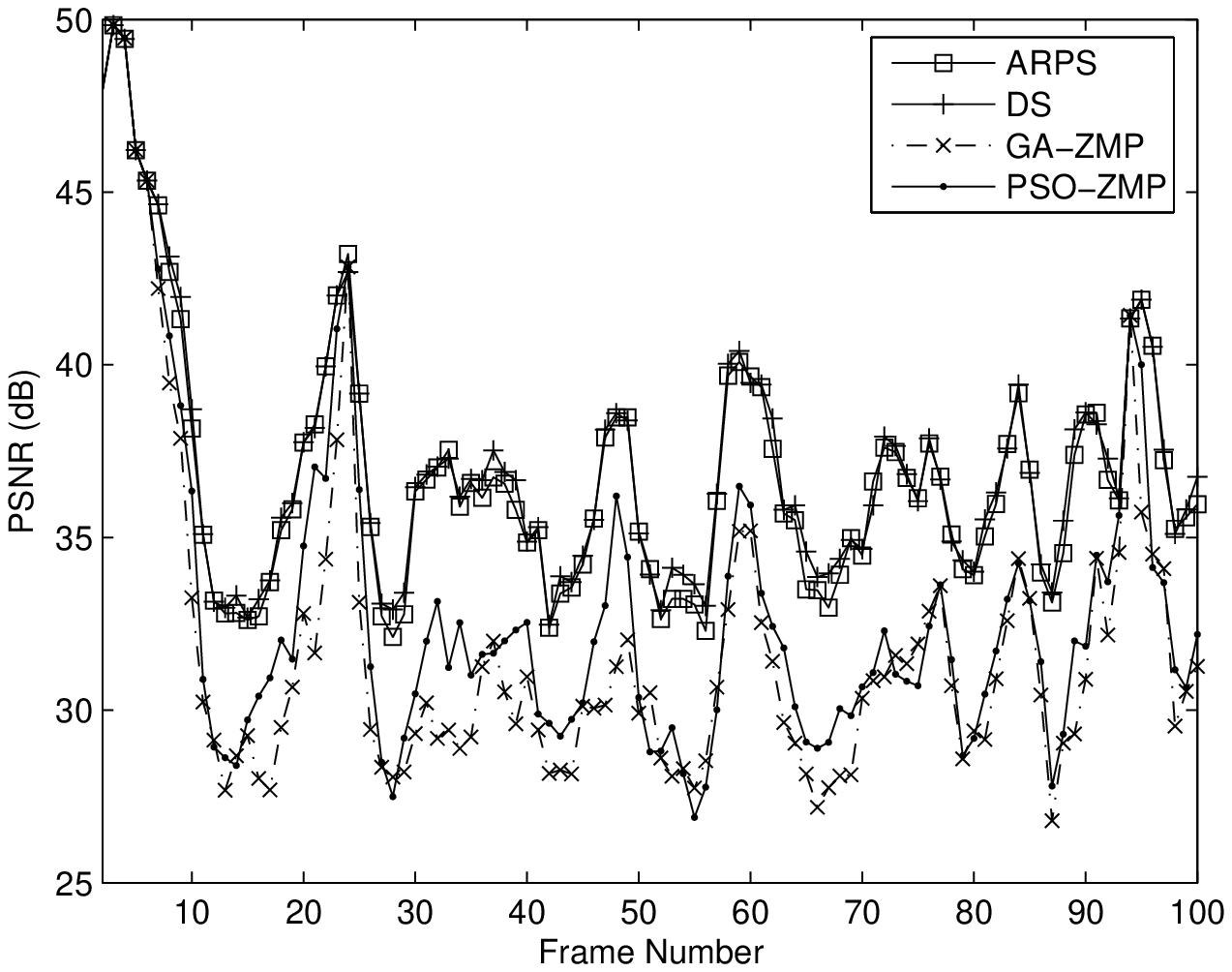}
\label{suzie.psnr}}} %
\caption{Simulation results on News and Silent} %
\label{fig_sim_2}
\end{figure*}

\section{Conclusion}
In this paper, we have proposed a fast block-based motion
estimation algorithm based on \textit{Particle Swarm
Optimization}(PSO) with novel particle initiation patterns.
Applied successfully in many functional and combinatorial
optimization problems, PSO is proved to have a relevant stronger
ability in global exploration. In addition, a \textit{zero-motion
prejudgment}(ZMP) routine is incorporated into the PSO BMA to
further reduce the computational cost of the algorithm. Simulation
results show that the PSO-ZMP BMA proposed requires less amount of
computation and achieves PSNR in a acceptable degree of drop.
while close and acceptable PSNR performance compared to widely
accepted ARPS and DS BMA. Moreover PSO just consumes a few lines
of codes due to its simplicity which makes the PSO-ZMP algorithm
attractive for hardware implementation.

In the future, variants of PSO might be applied to strengthen the
global searching ability and the accelerate the convergence speed.
And, to speed up the search and avoid being trapped in local
minima, a multiresolution procedure may be used.

\section*{Acknowledgment}
The authors would like to thank Qian Wu for helping us make the
nice search pattern figures and Yuxuan Wang for the invaluable
discussion and proofreading.

\bibliographystyle{abbrv}

\end{document}